\documentclass[pre,showpacs,letterpaper,twocolumn,floatfix]{revtex4}

\usepackage{bm}
\usepackage{bbm}
\usepackage{amsmath}
\usepackage{amssymb}
\usepackage{graphicx}
\usepackage{units}
\usepackage{hyperref}
\usepackage{url}

\begin{document}
\title{Dynamic radiation force of acoustic waves on solid elastic spheres}
\author{Glauber T. Silva}
\email{glauber@tci.ufal.br}
\affiliation{Centro de Pesquisa em Matem\'atica Computacional, Universidade Federal de Alagoas, Macei\'o, AL, Brasil, 57072-970}
\date{\today}

\begin{abstract}
The present study concerns the dynamic radiation force on solid elastic spheres exerted by a  plane wave with two frequencies (bichromatic wave) considering the nonlinearity of the fluid.
Our approach is based on solving the wave scattering for the sphere in the quasilinear approximation within the preshock wave range.
The dynamic radiation force is then obtained by integrating the component of the momentum flux tensor at the difference of the primary frequencies over the boundary of the sphere.
Results reveal that effects of the nonlinearity of the fluid plays a major role in dynamic radiation force leading it to a parametric amplification regime.
The developed theory is used to calculate the dynamic radiation force on three different solid spheres (aluminium, silver, and tungsten).
Resonances are observed in the spectrum of the force on the spheres.
They have larger amplitude and better shape than resonances present in static radiation force.
\end{abstract}
\maketitle

\section{Introduction}
Among nonlinear effects of acoustic wave propagation in fluids is worth noticeable radiation pressure (force)~\cite{borgnis53b} and streaming~\cite{eckart48a}.
These phenomena come together as an incident wave encounters an obstacle.
They are physically different but may combine producing an effective force upon the obstacle.
Usually these phenomena cannot be separated in actual fluids~\cite{PhysRevE.64.026311}.
When thermoviscous effects are negligible, streaming can be ruled out.
With respect to time dependence, radiation force is either static or dynamic.
Static radiation force is a time averaged quantity produced on a target by a monochromatic acoustic wave.
The dynamic counter-part is an oscillatory force exerted on a target by a bichromatic acoustic wave.
Here we consider that this force oscillates in time at the difference frequency of the wave, i.e., the difference between the primary wave frequencies.

The utilization of dynamic radiation force in science and technology has broadened.
It has been applied for measuring ultrasound power of ultrasonic transducers~\cite{greenspan78a}, inducing oscillation in bubbles and liquid drops~\cite{marston80a,marston:1233}, and exciting modes in capillary bridges~\cite{morse1996}.
Dynamic radiation force has become the underlying principle in some elastography techniques such as shear wave elasticity imaging~\cite{sarvazyan98a} and vibro-acoustography~\cite{fatemi98a}. 
Using vibrometry, the viscoelasticity of transparent materials can be determined by locally measuring the vibration of a embedded sphere induced by dynamic radiation force~\cite{chen02a}.

In a pioneer work, King~\cite{king34a} calculated the static radiation force exerted on a rigid sphere by a plane traveling wave in a lossless fluid.
Study on the static force of compressible and elastic spheres followed up~\cite{yosioka55a,hasegawa69a}.
It was promptly noticed that the static radiation force exhibits fluctuations related to the mechanical properties of the sphere.
Effects of viscosity and thermal conductivity of the host fluid on the static radiation force over a rigid sphere were also theoretically accounted~\cite{doinikov96a}.
For elastic solid media, a theoretical formulation of radiation force problem can be found in Ref.~\cite{potapov:441}.

The influence of the nonlinearity of the fluid on static radiation force is negligible at least for lossless fluids where the nonlinear parameter $B/A<100$~\cite{rooney73a,beissner95a}.
The dynamic radiation force exerted on a solid sphere by a plane traveling wave has been theoretically investigated in Ref.~\cite{silva2005a}.
Experimental validation of the theory is found in Ref.~\cite{chen05a}.
In both works, the difference frequency was assumed to be very narrow.
On that account the magnitude of dynamic and static radiation forces are alike.
Dynamic radiation force exerted on an object is directly caused by the pressure wave at the difference frequency acting on the object.
It is known that the wave at the difference frequency might be subjected to parametric amplification~\cite{westervelt63}.
This suggests that dynamic radiation force itself should achieve a regime of parametric amplification.
Silva~\textit{et al.}~\cite{silva:234301} verified that the dynamic radiation force on an acrylic sphere can be greatly amplified as the difference frequency is increased.

The aim of this work is to analyze the dynamic radiation force exerted on a solid elastic sphere in the parametric regime.
This analysis should be carried considering the nonlinearity of the fluid in the problem.
To the best of the author's knowledge this problem has not yet been treated.
To accomplish this task we have to solve the acoustic scattering for the sphere in at least the quasilinear approximation.
In what follows we present, based on regular perturbation technique, the wave equations in the quasilinear approximation for lossless fluids.
The solution for a bichromatic plane traveling wave is derived from these equations within the preshock wave range.
Using the Mie scattering theory we obtain a description for the scattered waves by the elastic sphere.
The yielded result is used in an vector integral formula obtained for the dynamic radiation force.
We apply the developed theory to analyze the dynamic radiation force on three different elastic spheres, namely, aluminium, silver, and tungsten.
A connection of dynamic radiation force and the Resonance Scattering Theory (RST) is established.
The spectrum of the dynamic radiation force for the spheres is numerically evaluated.
Resonances are observed in the spectrum of the force.
They depend in a unique manner on the mechanical and elastic parameters of the sphere.
Thus, the resonances in the spectrum could be used for characterization of materials.
Furthermore, their magnitude are up to $\unit[81]{dB}$ higher than those found in static radiation force.

\section{Nonlinear wave propagation and scattering}
Consider a homogeneous and isotropic fluid with adiabatic speed of sound $c_0$, in which thermal conductivity and viscosity are neglected. 
The fluid has infinite extent and is characterized by the following acoustic fields: pressure $p$, density $\rho$, and particle velocity $\mathbf{v}=-\nabla \phi$.
The function $\phi$ is the velocity potential and $\nabla$ is the gradient operator. 
These fields are function of the position vector $\mathbf{r}$ and time $t$.
At rest, these quantities assume constant values
$p=p_{0}$, $\rho = \rho_{0}$, and $\mathbf{v}=0$.
The acoustic fields are governed by the dynamic equations of ideal fluids.
By using the regular perturbation technique, one can expand velocity potential in terms of the Mach number $\varepsilon\ll 1$ as
$\phi = \varepsilon \phi^{(1)} + \varepsilon^2 \phi^{(2)}+O(\varepsilon^3),$
where the super-indexes indicate the order of the potential functions.
Any analysis of radiation force has to be done considering at least the quasilinear approximation, i.e., second-order acoustic fields.

To obtain the dynamic radiation force over an object, we have to solve the scattering of a bichromatic wave for the object.
Consider a bichromatic traveling plane wave of finite-amplitude with primary frequencies $\omega_1$ and $\omega_{2}$ propagating along the $z$-axis. 
The wave is formed by a sinusoidal excitation at $z=0$.
Due to nonlinear nature of wave propagation, a third wave at the difference frequency $\omega_{21} = \omega_2 - \omega_1$ arises in the fluid.
Suppose the bichromatic wave hits a sphere of radius $a$ placed in the $z$-axis at the distance $z_0$ from the acoustic source.
The elastic sphere has density $\rho_1$, compressional and shear speed of sound denoted, respectively, by $c_c$ and $c_s$.
To simplify our analysis we consider $k_{21} z_0 \gg 1$ and $a \ll z_0$.
Let us denote the complex amplitudes of the potential functions (incident plus scattered) at the frequencies $\omega_1$, $\omega_2$, and $\omega_{21}$ by, respectively, $\hat{\phi}_1$, $\hat{\phi}_2$, and $\hat{\phi}_{21}$.
The total potential for scattering problem can be found using the Mie scattering theory.
Accordingly, the velocity potentials in terms of spherical waves in the spherical coordinates $(r,\theta,\varphi)$ is given by~\cite{silva:234301}
\begin{eqnarray}
\nonumber
\hat{\phi}_m &=& A_m\text{Re}\sum_{n=0}^{+\infty}(2n+1) i^n [ h^{(2)}_n(k_m r)\\ 
&+& S_n(x_m) h^{(1)}_n(k_m r) ]P_n(\cos \theta),\quad m=1,2,
\label{phi_m_modal}\\
\nonumber
\hat{\phi}_{21} &=& 
A_{21}\text{Re}\sum_{n=0}^{+\infty}(2n+1) i^n [ h^{(2)}_n(k_{21} r) \\
&+& S_n(x_{21}) h^{(1)}_n(k_{21} r) ]P_n(\cos \theta),
\label{phi_21_modal} %
\end{eqnarray}
where $A_m=c_0e^{-i k_m z_0}/(2k_m)$, $A_{21}=\gamma c_0 z_0 e^{-i k_{21} z_0}/4$, $i$ is the imaginary unit with $k_m=\omega_m/c_0$ and $k_{21}=\omega_{21}/c_0$.
The quantity $\gamma = 1 + B/A$, where $B/A$ being the so-called nonlinear parameter of the fluid.
The functions $h^{(1)}_n$ and $h_n^{(2)}$ are, respectively, the first- and second-kind spherical Hankel functions of $n$th-order, while $P_n$ is the Legendre polynomial of $n$th-order.

The quantity $S_n$ is the modal scattering function.
To find this function we apply the solid elastic-fluid boundary conditions on the sphere surface.
These conditions yield a linear system whose solution is
\begin{equation}
\label{s-matrix}
S_n =  \det\left[ 
\begin{matrix}
a_{11} & d_{12} & d_{13}\\
a_{21} & d_{22} & d_{23}\\
0 & d_{32} & d_{33}
\end{matrix}
\right]
\det \mbox{}^{-1} \left[
\begin{matrix}
d_{11} & d_{12} & d_{13}\\
d_{21} & d_{22} & d_{23}\\
0 & d_{32} & d_{33}
\end{matrix}
\right].
\end{equation}
The matrix elements of this equation are found in Ref.~\cite{gaunaurd1983}.

\section{Theory of dynamic radiation force}
Let the Fourier transform of a function $g(t)$ be denoted by $\mathcal{F}[g]$.
The inverse transform is represented by $\mathcal{F}^{-1}$.
The dynamic radiation force exerted by a bichromatic wave on an object is given by~\cite{silva2005a}
\begin{equation}
\label{f21_t}
\mathbf{f}_{21}(t) = \mathcal{F}^{-1}\left[\tilde{\mathbf{f}}_{21}\right],
\end{equation}
where
\begin{align}
\nonumber
\tilde{\mathbf{f}}_{21} = &-\iint_{S_0} \mathcal{F}\biggl[p^{(2)}\mathbf{n} + \rho_0 \mathbf{v}^{(1)}\mathbf{v}^{(1)}\cdot\mathbf{n}\biggr]_{\omega_{21}} d S\\
&- i\varepsilon\omega_{21}\rho_0\mathcal{F}\left[
\iint_{S(t)}\phi^{(1)}\mathbf{n}d S\right]_{\omega_{21}}.
\label{f21}
\end{align}
The quantity $\rho_0 \mathbf{v}^{(1)}\mathbf{v}^{(1)}$ is the Reynolds' stress tensor, while $p^{(2)}$ is the second-order pressure given by
\begin{equation}
\label{2nd_pressure}
p^{(2)} = \varepsilon^2 \rho_0 \left[\frac{1}{2c_0^2} \left(\frac{\partial \phi^{(1)}}{\partial t}\right)^2 -\frac{\|\nabla \phi^{(1)}\|^2}{2}  +  \frac{\partial \phi^{(2)}}{\partial t}\right]
\end{equation}
We restrict our analysis to waves in which $\omega_{21}<\omega_1$, otherwise we may have superposition of the dynamic radiation force at $\omega_{21}$ and other forces caused by the waves at the primary frequencies.

The second integral in the right-hand side of Eq.~(\ref{f21}) depends on the variation of the object boundary caused by the incident wave.
The evaluation of this integral for a moving object is a hard task because this is a self-coupled problem.
We have to calculate the force on the object in first-order by integrating the linear pressure on the boundary of the object.
However, the pressure may deform or move the boundary of the object.
We shall show an approximate calculation of this term for a rigid spherical target.

Consider a rigid sphere of radius $a$ and density $\rho_1$ suspended in a fluid.
The origin of the coordinate system coincide to the center of the sphere at rest.
We assume the sphere oscillates only along the $z$-axis.
The variation $\Delta z$ in the first-order approximation of the normal vector of the sphere is given by
$$\mathbf{n}= \mathbf{e}_r -\Delta z\mathbf{e}_z + O(\varepsilon^2\mathbf{e}_z),$$
where $\mathbf{e}_r$ is the radial unit vector and $\mathbf{e}_z$ is the unit vector along the $z$ direction.
We have $\Delta z = \delta z_1 e^{-i\omega_1 t} + \delta z_2 e^{-i\omega_2 t}$, $\delta z_1$ and $\delta z_2$ are the amplitudes of oscillation of the sphere.
The second integral of Eq.~(\ref{f21}) might be approximated to
\begin{equation}
\label{int_p}
\mathcal{F}\left[\iint_{S(t)}\phi^{(1)}\mathbf{n}d S\right]_{\omega_{21}}
\simeq \frac{\mathbf{e}_z}{a}\iint_{S_0}\mathcal{F}\left[\phi^{(1)}\Delta z \right	]_{\omega_{21}}d S.
\end{equation}
The amplitudes of oscillation are given in terms of the mechanical impedances of the sphere $Z_1$ and $Z_2$ at the frequencies $\omega_1$ and $\omega_2$ as~\cite{oestreicher1951}	
\begin{equation}
\label{z_m}
\delta z_m = \frac{i}{\omega_m Z_m} \iint_{S_0} p_m^{(1)} d S, \quad m=1,2;
\end{equation}
where
$p^{(1)}_m=\varepsilon \rho_0 \partial \phi^{(1)}_m/\partial t$ is the linear pressure and
\begin{equation}
\label{impedance}
Z_m = i \frac{4}{3}\pi a^3 \omega_m \left[\rho_0  \frac{2+x_m^2 - i x_m^3}{4+x_m^4}+\rho_1\right], \quad m=1,2.
\end{equation}
The quantity $x_m=\omega_m a/c_0$ is the size factor of the sphere. 
Equations (\ref{int_p}) and (\ref{z_m}) give us an approximation to calculate the first term of the amplitudes of the dynamic radiation force in Eq.~(\ref{f21}).

The dynamic radiation force is calculated as follows.
With the obtained first- and second-order potential fields in Eqs.~(\ref{phi_m_modal}) and (\ref{phi_21_modal}), the first term in the right-hand side of Eq.~(\ref{f21}) is calculated performing the surface integral.
The second integral of this equation can be evaluated using Eqs.~(\ref{int_p}), (\ref{z_m}), and (\ref{impedance}).
Accordingly, the dynamic radiation force on the sphere is given by
\begin{equation}
\label{force12}
\mathbf{f}_{21}(t) = \pi a^2 E_0 \hat{Y}_{21} e^{-i (\omega_{21} t - k_{21}z_0)} \mathbf{e}_z,
\end{equation}
where $E_0 = \varepsilon^2 \rho_0 c_0^2/2$ is the energy density at the acoustic source and $\hat{Y}_{21}$ is the radiation force function.
This function might be decomposed as $\hat{Y}_{21} = \hat{Y}_{a}+\hat{Y}_b+\hat{Y}_c$, where
\begin{eqnarray}
\nonumber
\hat{Y}_a &=& -i \pi a^2 \rho_0 c_0  \left(Z_1^{-1*}+Z_2^{-1}\right) \frac{ x_{21} }{x_1 x_2} R_0(x_2)R^*_0(x_1),\\
\label{Ya}\\
\label{Yb}
\hat{Y}_b &=& -\frac{i \gamma z_0}{a}  {x}_{21}R_1(x_{21}),
\end{eqnarray}
and
\begin{widetext}
\begin{eqnarray}
\nonumber
\hat{Y}_c &=& - \sum_{n=0}^{+\infty} \frac{n+1}{ {x}_2  {x}_1}
\biggl\{ \left[ {x}_2  {x}_1 - n(n+2)\right]\left[R_{n}(x_2) R_{n+1}^*(x_1)
+ R_{n+1}(x_2) R_{n}^*(x_1)\right]- n [ {x}_1 R_{n}(x_2) R\mbox{}'^{*}_{n+1}(x_1)\\
\nonumber
&\mbox{}&
+~ {x}_2 R'_{n+1}(x_2) R_{n}^{*}(x_1)]+ (n+2)\left[ {x}_2 R'_{n}(x_2) R_{n+1}^{*}(x_1)
+  {x}_1 R_{n+1}(x_2) R\mbox{}'^{*}_{n}(x_1)\right]
+  {x}_2  {x}_1 [R'_{n}(x_2) R\mbox{}'^{*}_{n+1}(x_1)\\
&\mbox{}& 
+~ R'_{n+1}(x_2) R\mbox{}'^{*}_{n}(x_2)]\biggr\},
\label{Yc}
\end{eqnarray}
\end{widetext}
where $R_n(x) = i^n [ h^{(2)}_n(x) + S_n(x) h^{(1)}_n(x)]$ and
$R'_n(x) = i^n [ {h^{(2)}_n}'(x) + S_n(x) {h^{(1)}_n}'(x)]$.
The symbol $'$ means the derivative of the function with respect to its argument.
The function $\hat{Y}_{21}$ depends on the size factors $x_2$ and $x_1$ with the restriction $x_{21}<x_1<x_2$.

The function $\hat{Y}_a$ is proportional to the inverse of the radiation impedance of the sphere at the fundamental frequencies $\omega_1$ and $\omega_2$.
This term is divergent when the size factors $x_1$ and $x_2$ approach to zero.
In the case $x_1,x_2>1$ we may disregard the contribution of $\hat{Y}_a$ to the radiation force function.
Furthermore, 
$
\underset{x_{21}\rightarrow 0}{\lim}  \hat{Y}_a = 0.
$
The contribution of function $\hat{Y}_b$ brings up the regime of parametric amplification in the dynamic radiation force.
It has a dominant role whenever the size factor $x_{21}$ is not a small quantity.
It also depends linearly on the nonlinear parameter $\gamma$ and the ratio $z_0/a$.
This function has the following asymptotic behavior
$ |\hat{Y}_b| \underset{x_{21}\rightarrow+\infty}{\sim} 2z_0\gamma a^{-1} \left(1+x_{21}^{-1}\right).$
The amplification of the radiation force with $z_0$ does not increase indefinitely.
On that account two aspects should be asserted.
The distance $z_0$ should be within the preshock wave range otherwise Eqs.~(\ref{phi_m_modal}) and (\ref{phi_21_modal}) are no longer valid.
Diffraction and fluid viscosity may alter the dependence of radiation force with the distance $z_0$ (see Ref.~\cite{tjotta1981}).
The function $\hat{Y}_c$ has a prominent contribution to the dynamic radiation force whenever $x_{mn}\ll 1$.
When $x_{21}=0$ the incident wave becomes monochromatic and the dynamic radiation force in Eq.~(\ref{force12}) reduces to its static counter-part as obtained in Ref.~\cite{hasegawa69a}.

Hereafter, we focus on the situation of which $x_{21}$ is not a very small quantity and $x_1,x_2>1$.
Therefore, the radiation force function becomes $\hat{Y}_{21}=\hat{Y}_b$.
In that case, the integral formula of dynamic radiation force in Eq.~(\ref{f21}) is reduced to
\begin{equation}
\label{f21_simpler}
\mathbf{f}_{21}(t) = -\iint_{S_0} p_{21}^{(2)}(t)\mathbf{n}dS,
\end{equation}
where $p^{(2)}_{21} = \varepsilon^2 \rho_0 \partial \phi^{(2)}_{21}/\partial t$ with $\phi^{(2)}_{21}$ being the second-order velocity potential at the difference frequency.
Notice the case for which $x_{21} \ll 1$ has been analyzed elsewhere~\cite{silva2005a}.

\section{Resonance scattering theory and dynamic radiation force}
In our previous analysis, we have found that the dynamic radiation force function reduces to  $\hat{Y}_{21}=\hat{Y}_b$.
According to the Resonance Scattering Theory (RST), the response of an elastic sphere to a plane wave excitation can be decomposed into a modal rigid body background interfering with resonance contributions~\cite{gaunaurd1983}.
The resonances appear due to the eigenvibration modes of some types of surface waves (Rayleigh and whispering gallery waves) present on the sphere as the result of the plane wave excitation.

To isolate the resonance contributions to the dynamic radiation force, we follow the approach developed in Ref.~\cite{rhee:3401}.
Accordingly, the resonance scattering function is given by
\begin{equation}
S^{(\text{res})}_n = \frac{S_n}{S_n^{(\text{r})}} - 1,
\end{equation}
where $S_n^{(\text{r})}(x) = h_1^{(2)}\mbox{}'(x)/h_1^{(1)}\mbox{}'(x)$ is the modal scattering function of a rigid sphere.
We thus define the resonance radiation force function of an elastic sphere as
\begin{equation}
\hat{Y}_{21}^{(\text{res})} = x_{21}S_1^{(\text{res})}(x_{21})h_1^{(1)}(x_{21}),
\label{Yres}
\end{equation}
for which we dropped the factor $\gamma z_0/a$ because it is not relevant to resonance analysis.
The resonance radiation force function is proportional to the backscattered wave $(\theta=\pi)$ of mode $n=1$ (the dipole mode).
Furthermore, this function carries an unique information of the sphere in terms of its mechanical parameters, including the elastic constants of the material.

\section{Numerical results}
We evaluate the radiation force function for three different solid spheres, which are made aluminium, silver, and tungsten. 
The physical parameters under consideration of these materials are summarized in Table~\ref{tab:parameters}.
The fluid medium is water whose parameters are $\rho_0=\unit[1000]{Kg/m^3}$, $c_0=\unit[1500]{m/s}$, and $\gamma=6$.
\begin{table}[h]
\caption{Physical parameters for the spheres after Ref.~\cite{anson:1618}.}
\label{tab:parameters}
\begin{center}
\begin{tabular}{lccc}
\hline
\hline
&&\multicolumn{2}{c}{Speed of sound}\\
\cline{3-4}
Material 
& Density
& Compressional
& Shear\\
&
[Kg/m$^3$]&
[m/s]&
[m/s]\\
\hline
Aluminium (Al) & ~2700 & 6374 & 3111 \\
Silver (Ag)& 10500 & 3704 & 1698 \\
Tungsten (W) & 19250 & 5221 & 2887 \\
\hline
\hline
\end{tabular}
\end{center}
\end{table}

\begin{figure}
\begin{center}
	\includegraphics[width=\linewidth]{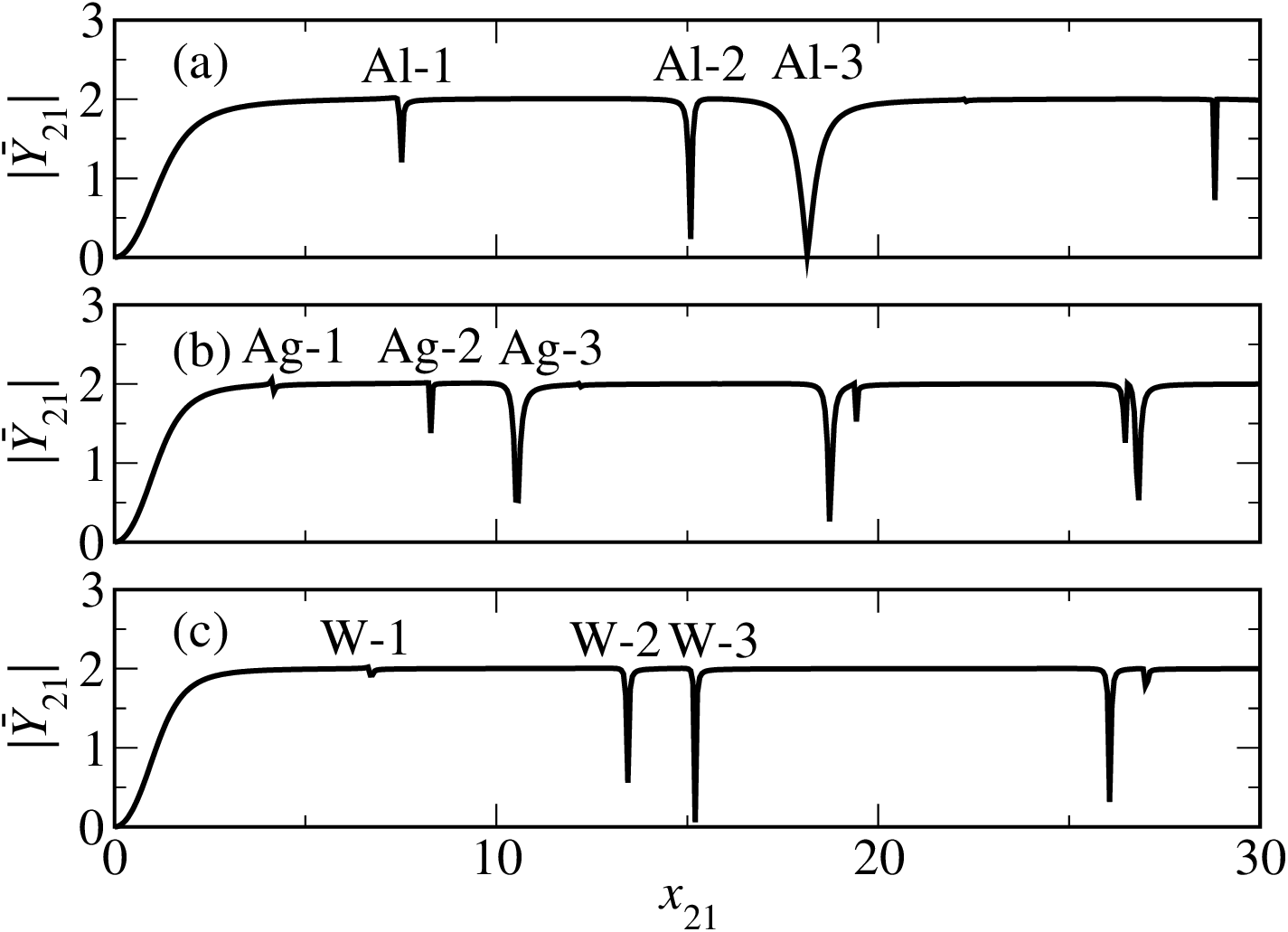}
\end{center}
\caption{Magnitude of the radiation force function $\bar{Y}_{21}$. 
The labels Al-$n$, Ag-$n$, and W-$n$ indicate the $n$th resonance in the radiation force function. (a) Aluminium. (b) Silver. (c) Tungsten.}
\label{fig:rad_force_mag}
\end{figure}

In Fig.~\ref{fig:rad_force_mag}, we plot the spectrum of the function $\bar{Y}_{21} = a(\gamma z_0)^{-1}\hat{Y}_{21}$ for the three spheres.
This function varies within the range $0<x_{21}<30$ incremented by steps of $0.01$.
We also consider $x_1,x_2>1$, therefore, the contributions from $\hat{Y}_a$ and $\hat{Y}_c$ to the radiation force function are negligible.
The region $0<x_{21}<5$ corresponds to the crossover between the static radiation force $(x_{21}=0)$ and the fully developed dynamic radiation force $x_{21}>5$.
To see how larger is the magnitude of the dynamic radiation force in this region compared to that of the static counter-part consider $z_0/a = 1000$.
The magnitude of the static radiation force function for all spheres is about the unit~\cite{anson:1618}, while that for the dynamic radiation force function in $x_{21}>5$ is around $12000$.
Therefore, the magnitude of the fully developed dynamic radiation force is $\unit[81]{dB}$ higher than that of the static radiation force.
The spectrum of dynamic radiation force function follows the behavior of the radiation force function for a rigid sphere in which $c_c,c_s,\rho_1\rightarrow +\infty$, except for the the appearance of well defined resonances.
The labels Al-$n$, Ag-$n$, and W-$n$ indicate the $n$th resonance in the radiation force function for, respectively, the aluminium, the silver, and the tungsten spheres.
Only the first three resonances are highlighted in the graphs.

\begin{figure}
\begin{center}
	\includegraphics[width=\linewidth]{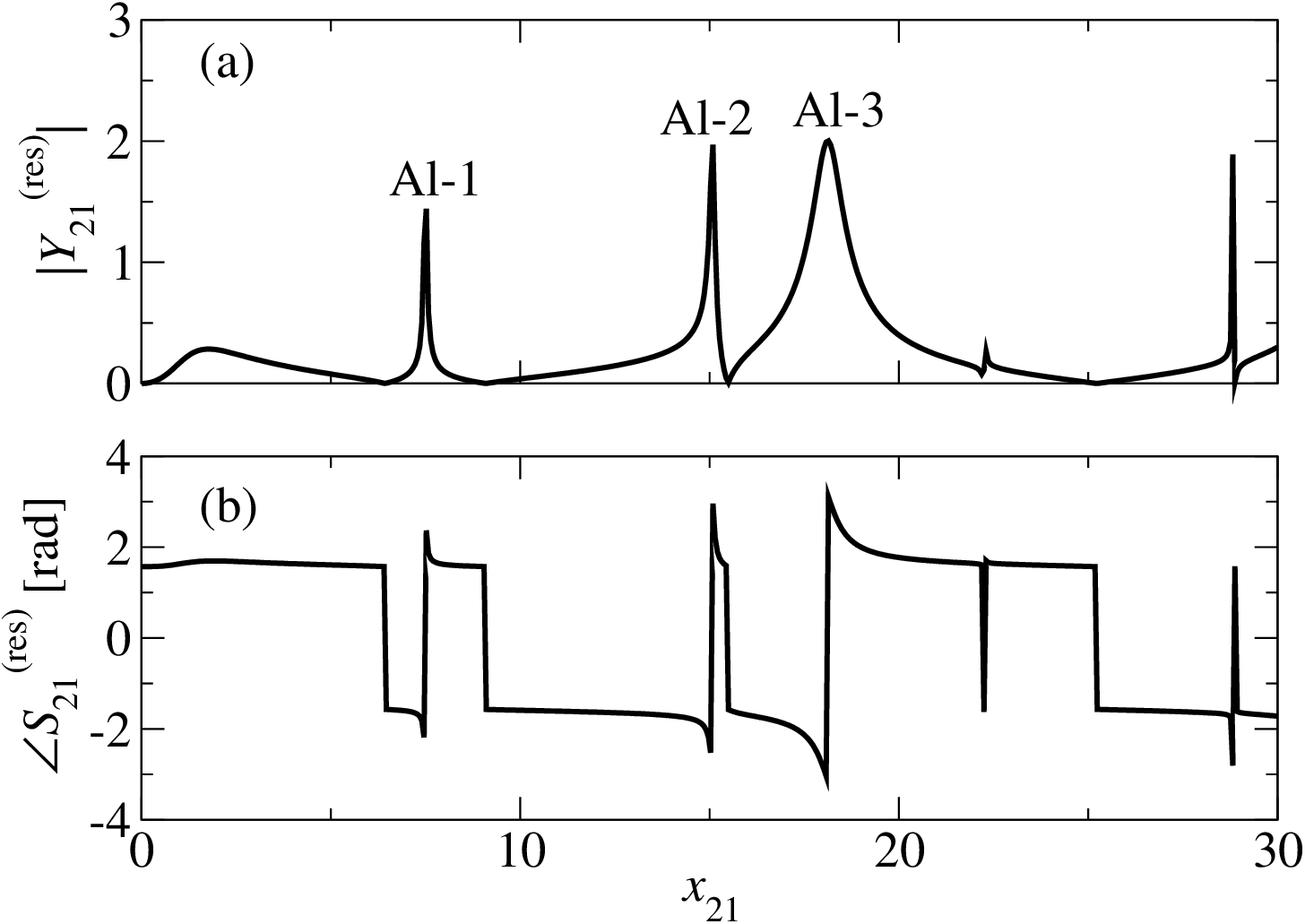}
\end{center}
\caption{The spectrum of the resonance radiation force of the aluminium sphere. 
(a) Magnitude. (b) Phase of the resonance scattering function $S^{(\text{res})}_{21}$.}
\label{fig:resonance_aluminium}
\end{figure}

The spectrum of the resonance radiation force function for the aluminium sphere is plotted in Fig.~\ref{fig:resonance_aluminium}.a.
We plot the phase of the corresponding resonance scattering function in Fig.~\ref{fig:resonance_aluminium}.b.
Whenever the spectrum undergoes through a resonance or an antiresonance (destructive interference of two adjacent peaks~\cite{ewins1986})
the phase shifts by $\pi$.
The almost constant behavior of the phase near to the first peak reveals that the peak is not related to a resonance of the sphere.

\begin{figure}
\begin{center}
	\includegraphics[width=\linewidth]{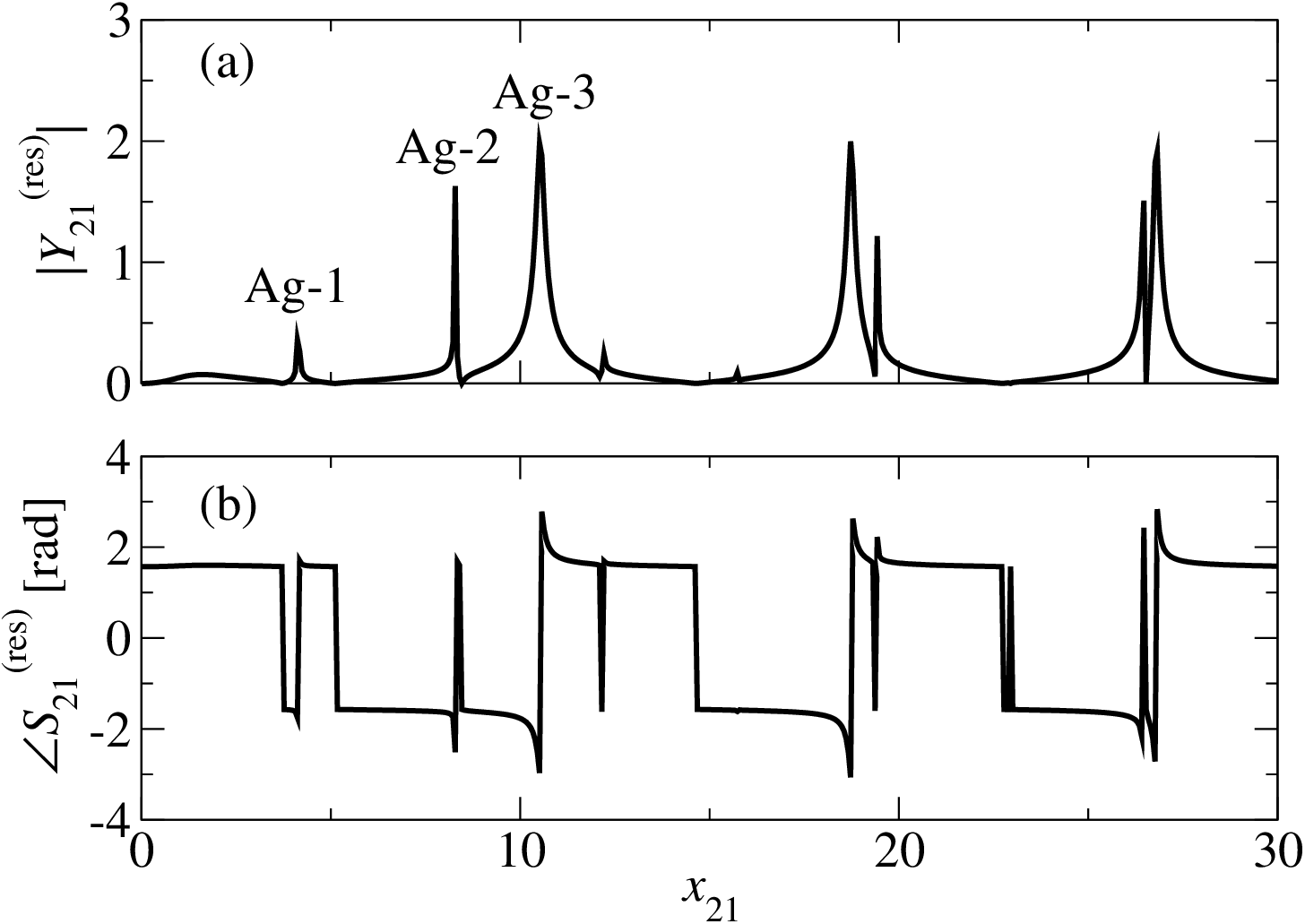}
\end{center}
\caption{The spectrum of the resonance radiation force function of the silver sphere. (a) Magnitude. (b) Phase of the resonance scattering function $S^{(\text{res})}_{21}$.}
\label{fig:resonance_silver}
\end{figure}

In Fig.~\ref{fig:resonance_silver}.a, we have the spectrum of the resonance radiation force function for the silver sphere.
The phase of the resonance scattering function is shown in Fig.~\ref{fig:resonance_silver}.b.
We have a similar behavior of the spectrum as observed for the aluminium sphere.

\begin{figure}[t]
\begin{center}
	\includegraphics[width=\linewidth]{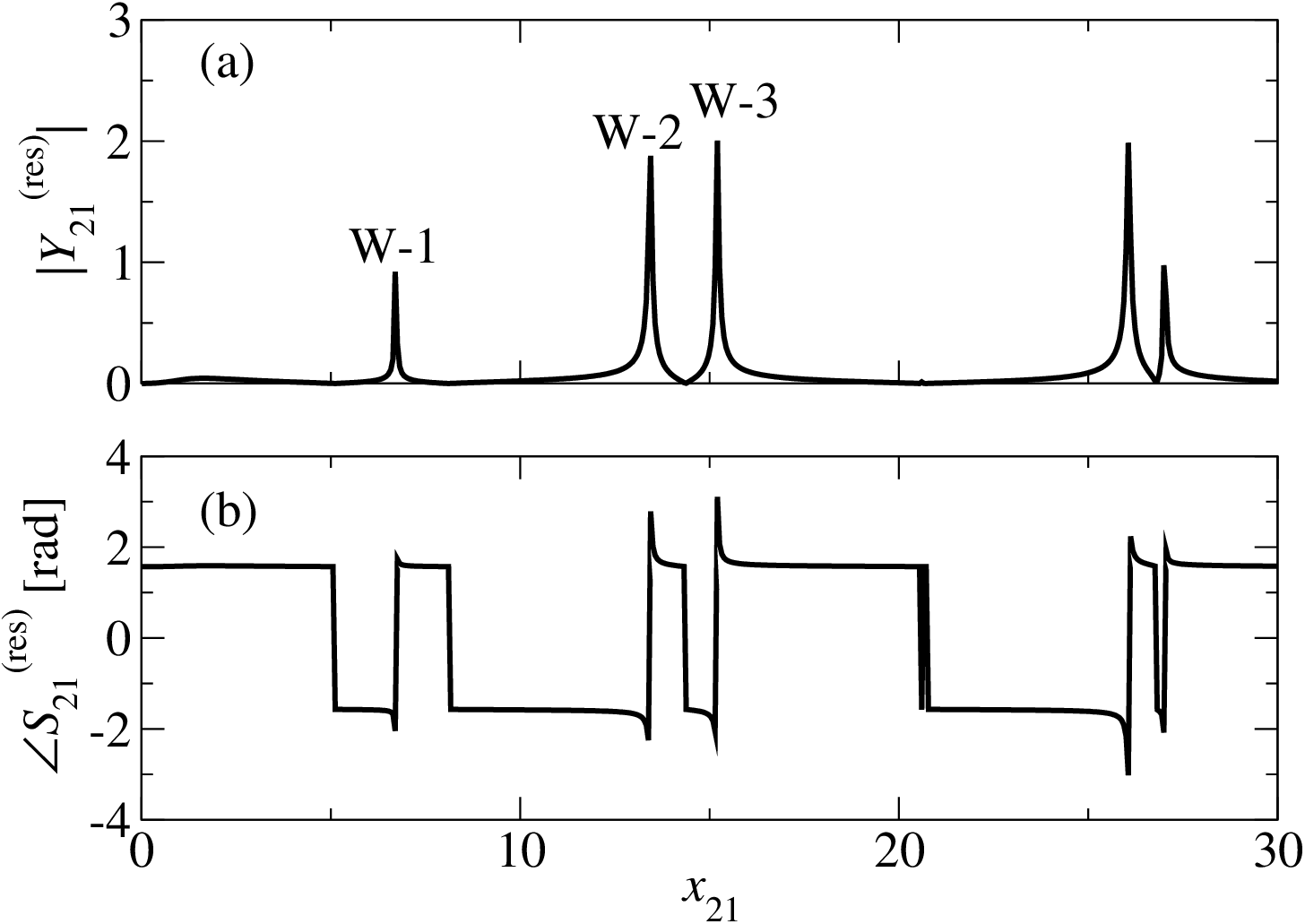}
\end{center}
\caption{The spectrum of the resonance radiation force function of the tungsten sphere. (a) Magnitude. (b) Phase of the resonance scattering function $S^{(\text{res})}_{21}$.}
\label{fig:resonance_tungsten}
\end{figure}

The spectrum of the resonance radiation force function for the tungsten sphere is presented in Fig.~\ref{fig:resonance_tungsten}.
The behavior of the resonances as well as the phase of $S^\text{(res)}_1$ follow those of the aluminium and the silver spheres.

\section{Discussion and conclusion}
We have carried out a theoretical analysis of dynamic radiation force caused by bichromatic waves on a target taking into account  the nonlinearty of the fluid medium.
In the analysis, the object target must be within the preshock wave range.
To obtain the force exerted on an solid elastic sphere we had to solve the scattering problem in the quasilinear approximation using the partial wave expansions given in Eqs.~(\ref{phi_m_modal}) and (\ref{phi_21_modal}).
The incident and scattered fields were used in Eq.~(\ref{f21}) yielding the radiation force.
It was demonstrated that dynamic radiation force is caused by parametric amplification of the difference frequency wave.
This is experimentally supported by Ref.~\cite{silva:234301}.
Further, the force is proportional to the nonlinear parameter of the fluid $\gamma$. 
We therefore presented a much simpler formula for dynamic radiation force in Eq.~(\ref{f21_simpler}).

The obtained result of the dynamic radiation function in Eq.~(\ref{Yb}) permitted a straight connection with this force and the Resonance Scattering Theory (RST).
To the best of the author's knowledge such connection has not yet been explored before, neither theoretically nor experimentally.
The spectrum of the dynamic radiation force as a function $x_{21}$ exhibited well defined resonances, which have larger amplitude (about $\unit[81]{dB}$ higher) and are better shaped than those present in static radiation force~\cite{anson:1618}.
The resonances are caused by eigenvibrations in some surface waves on the sphere at the difference frequency $\omega_{21}$.
The RST allowed us to separate the resonance part of the radiation force function as given in Eq.~(\ref{Yres}).
We, thus, analyze the resonance spectrum for three different spheres made of aluminium, silver, and tungsten.
Furthermore, the study of dynamic radiation force on spherical shells can be readily done by using the present theory.
The location, amplitude, and width of these resonances depend upon the mechanical and the elastic properties of the sphere.
Hence, the spectrum information could be used as the material signature in a noncontact acoustic spectroscopy technique. 
In this method, the induced vibration by the dynamic radiation force on a specimen could be measured by a laser vibrometer yielding the spectrum of the force~\cite{chen05a}.
From the spectrum it is possible to extract the elastic properties of the specimen in a similar manner as done in the Resonant Ultrasound Spectroscopy~\cite{maynard:26}.

We have not considered thermoviscous effects in the fluid.
By neglecting such effects we ruled out acoustic streaming in the boundary of the sphere.
Usually streaming has the tendency of altering the total effective force acting on a target.
Moreover, the dependency of dynamic radiation force with temperature was not accomplished here.
Variations in the temperature may change the location and width of the resonances in the spectrum of the dynamic radiation force.

In conclusion, this paper presented a comprehensive theoretical analysis of the dynamic radiation force on solid spheres.
The spectrum of this force carries useful information of the sphere related to its elastic and mechanical properties.

\begin{acknowledgments}
This work was supported by FAPEAL/CNPq (Brazilian Agencies).
\end{acknowledgments}



\end{document}